# Fractal properties of isolines at varying altitude reveal different dominant geological processes on Earth

Andrea Baldassarri[1], Marco Montuori[2], Olga Prieto-Ballesteros[3], and Susanna C. Manrubia[3]

[1]*Dipartimento di Fisica, Università "La Sapienza", P.le A. Moro 2, 00185 Roma, Italy*
[2]*CNR-INFM, Pl.Aldo Moro 7, 00185 Roma, Italy*
[3]*Centro de Astrobiología, CSIC-INTA, Ctra. de Ajalvir km. 4, 28850 Torrejón de Ardoz, Madrid, Spain*

**Abstract**

Geometrical properties of landscapes result from the geological processes that have acted through time. The quantitative analysis of natural relief represents an objective form of aiding in the visual interpretation of landscapes, as studies on coastlines, river networks, and global topography, have shown. Still, an open question is whether a clear relationship between the quantitative properties of landscapes and the dominant geomorphologic processes that originate them can be established. In this contribution, we show that the geometry of topographic isolines is an appropriate observable to help disentangle such a relationship. A fractal analysis of terrestrial isolines yields a clear identification of trenches and abyssal plains, differentiates oceanic ridges from continental slopes and platforms, localizes coastlines and river systems, and isolates areas at high elevation (or latitude) subjected to the erosive action of ice. The study of the geometrical properties of the Lunar landscape supports the existence of a correspondence between principal geomorphic processes and landforms. Our analysis can be easily applied to other planetary bodies.

# 1. Introduction

There exists an overwhelming diversity of landscapes on Earth. A cornerstone of modern geomorphology came with the realization that all the different features of the terrestrial surface result from the accumulated effect of current geological agents [*Lyell,* 1830]. This principle established for the first time a qualitative relationship between pattern and process in geology. More than one century later, with the application of fractal geometry to natural systems, the characteristic, rough shapes exhibited by landscapes were identified in a first approximation as *self-similar*, triggering in this way the research on mechanistic and theoretical models aimed at identifying the underlying constructive rules responsible for the appearance of scale-invariant patterns. Computational analysis of the geometry of coastlines [*Mandelbrot,* 1967; 1983; *Boffetta et al.,* 2007] and river networks [*Mandelbrot,* 1967; *Hack,* 1957; *Rinaldo et al.,* 1993; *Rodríguez-Iturbe and Rinaldo,* 1997; *Nieman et al.*, 2001] placed them amongst the first natural systems quantitatively characterized. The fact that both systems display quasi-universal properties prompted the search for simple models able to account for their quantitative features. The topology of river networks seems to stem from a principle of transport optimization [*Maritan et al.,* 1996] while the dominant 4/3 fractal dimension of coasts can be retrieved from a damping erosion model [*Sapoval et al.,* 2004]. In comparison, global topography has received less attention, maybe reflecting that, at odds with coastlines and rivers, terrestrial topography possesses a more complex scale invariance (is multifractal [*Gagnon et al.,* 2006]) and its geometry depends on specific features of the region analysed [*Dodds and Rothman*, 2000; *Ivanov,* 1994; *Gagnon et al.,* 2003]. In all cases, fractal geometry has proven useful in the description of natural landscapes, due to the characteristic and ubiquitous repetition of similar motifs at different spatial scales. At present, one of the goals of topography analysis is to single out characteristic features with well defined scale invariant properties (in the sense used above for river networks or coasts) as a first step towards distinguishing dominant physical process, and the subsequent design of simple models able to reproduce the average geometrical properties of that region.

This ambitious goal is limited by our understanding of the complex feedback loop between the action of geological processes and the surface features of Earth. Geological agents settle the characteristics of the topography but act in turn with different strength as a function of elevation [*Weissel et al.,* 1994], the latter being a variable itself on geological time scales. Two clear examples of this interrelationship are coastlines and young mountains. The former are mainly shaped through water erosion and can only occur at the contacts among oceans, emerged lands, and atmosphere. The characteristics of young, high mountains result from orogenic processes and the action of liquid and solid water. Differentiation between lowlands and highlands, where altitude is correlated with geomorphic features, occurs in other planetary bodies as well. Mars is divided into cratered highlands, covering most of the southern hemisphere, and resurfaced lowlands, in the northern hemisphere [*Aharonson et al.*, 2001]. The surface of Venus can be divided in its turn into three main regions according to elevation: highlands are characterized by volcanic and strongly tectonized structures, mesolands are dominated by large extensional fracture belts and several volcanic features, and lowlands are mainly shaped through volcanism [*Bazilevski et al.*, 1982; *Price and Suppe*, 1995]. Taking into account the clear differentiation in morphological structures as a function of their elevation, an analysis of the global topography based on measures that mix several heights could jumble different features. Hence, the use of isolines, defined as the set of points at a fixed elevation, appears as a suitable solution to undertake a quantification of topography [*Isichenko,* 1992; *Kondev et al.*, 2000]. The fractal dimension $D$, a measure of the degree of self-similarity of a curve, will be used to characterize in a first approximation the geometrical properties of isolines.

At present, comparative geomorphology is the everyday tool used by geologists to infer the geological processes responsible for the structures observed not only on Earth, but on any planetary body. This qualitative analysis of classical geomorphology can experience a boost with the use of technical advances yielding direct and precise measurements of topographic relief. An accurate and objective quantification of such data is of prime relevance to identify major geological processes on Earth and elsewhere [*Dodds and Rothman*, 2000; *Ivanov,* 1994; *Gagnon et al.,* 2006; *Maritan et al.*, 1996; *Turcotte*, 1997; *Wilson and Dominic*, 1998; *Sung and Chen,* 2004]. In this work, we take advantage of the large amount of topographic data available to obtain reliable, three dimensional models of planetary surfaces. At present, Earth, Moon, and Mars are well characterized, so their global topography is susceptible to being quantitatively studied. Our aim is to introduce a formal methodology that represents a first step towards the unsupervised identification of major topographic characteristics of landscapes. The method we propose is based on the measure of the *local* fractal dimension of isolines on the whole planetary surface. A statistical analysis of the results advances in establishing a correspondence between topographic features and intervening geological agents.

## 2. Isoline Analysis

Topographic data for Earth have been obtained from the SRTM30-plus set. The data consists of measured elevations over a grid of surface points. The resolution is of 30" (approximately 1km) for continental land, and of 2' (approximately 4km) for seafloor topography [SRTM data, 2006]. Topographic data for the Moon come from the Clementine Project [*Nozette et al.*, 1994; Clementine data, 1994] and have a resolution of 30'. Our analysis will be restricted to the Lunar region spanning 70S to 70N, since direct measures of altimetry from the Clementine project covered approximately that band (Smith et al., 1997).

Isolines for both bodies have been identified at intervals of 100m. To determine the points belonging to an isoline at level $h$, we select pairs of neighbouring measures in the grid with positions $p_1$ and $p_2$ and corresponding elevations $h_1$ and $h_2$, with the condition that $h_1 \geq h \geq h_2$. We assayed three different definitions of isolines: (i) both $p_1$ and $p_2$ belong to the isoline at level $h$; (ii) only $p_2$ (c.f. $p_1$) belongs to the isoline; (iii) a single isolated point with averaged position (coordinates) between $p_1$ and $p_2$ belongs to the isoline. Either definition yielded exactly the same quantitative results. An isoline is finally a set of $N$ points, with $N$ not larger than the number of measures in the original, finite data set. This procedure is analogous to the *functional box counting* introduced in previous studies [*Lovejoy and Schertzer,* 1991].

We have computed the fractal dimension of each isoline through the usual box-counting method, using square boxes of linear size $r$ and calculating at each scale the number of boxes needed to cover all of the points in the isoline. The isolines we are measuring are not fractal objects in a strict mathematical sense. The spatial scale of potential self-similarity is bound from below by the resolution of the available data and from above by the requirement of analysing a geologically homogeneous region. Data become too sparse at small scales, such that in practice there are more empty small boxes that would be found if the isoline could be described with arbitrarily small resolution. This causes a bending towards lower fractal dimensions (slope zero is reached below the grid size, when the number of full boxes becomes independent of scale). Further, we are carrying out a systematic, unsupervised analysis of a huge data set. This constrains the range by demanding a minimum and maximum linear scale valid for all situations. Eventually, interpolation to obtain the fractal dimension will be performed within a conservative range [$r_{min}$, $r_{max}$] with $r_{min}$ in the order of few km. and $r_{max}$ around several tens of km. As an example, in the terrestrial region between 60S and 60N, we work in angular coordinates corresponding to $r_{min}$ =0.06° and $r_{max}$ =0.6° (at terrestrial equator, 1° longitude spans about 111km). Similarly, for the Moon we use $r_{min}$ =0.5° and $r_{max}$ =5.0°. Measures of the fractal dimension of isolines are performed on a set of angular regions, hereafter denoted "cells" that fully cover the planetary surface. For the Earth we choose a cell size of 4° latitude x 4° longitude, and for the Moon 15° latitude x 15° longitude. Linear size of cells is chosen about ten times larger than $r_{max}$, once more in order to avoid size effects (too close to the cell size all boxes are full and the fractal dimension crosses-over to value 2). In each cell, we computed the fractal dimension for every isoline present. As a result, we get tens of thousands of measures, each one corresponding to a fractal dimension $D(lon,lat,h)$ where $h$ is the elevation of the portion of isoline contained in the cell centred at longitude *lon* and latitude *lat*.

Figure 1 shows the results of box-counting performed on six terrestrial cells at different elevations, as shown in the legend. The corresponding cells and isolines are plotted in Fig. 6(c) (two examples at elevation 2000m are shown, only one of them appears in 6(c)), where the geographical location is also given. As can be seen, all isolines are fairly self-similar in the range selected for interpolation (straight lines) and even at larger or smaller scales. The bending at small scales due to insufficient sampling is observed in all cases. Finally, there is a clear variation in fractal dimension as a function of elevation.

In our analyses, we have discarded fractal dimensions obtained from isolines with less than $N$=500 points for Earth and $N$=200 points for the Moon. The regression error in the fractal dimension $D$ never exceeded 4%. We tested the robustness of our results by repeating the whole numerical analysis in the following cases: (i) full resolution for continental land (30" instead of 2'); (ii) different minimum number of points per isoline ($N$=100, 1000, and 2000 for the Earth, and $N$=100 and 500 for the Moon); (iv) pruning of the isolines from local connected "islands" (for instance resulting from peaks of mountains or very small craters that could contribute, in the spatial range considered, as a dust of isolated points systematically affecting the measures); (v) changing the size of the tessellation of the planetary surfaces (cell size up to 60° latitude x 60° longitude) and the value of $r_{max}$ up to 6°. The values of the corresponding measured fractal dimensions are slightly affected quantitatively, but the main results do not change. For instance the curves of average fractal dimensions vs. elevations show a vertical global shift by about 0.1, though the presence of a qualitative signal which identifies major features from a statistical viewpoint is robust with respect to the enumerated modifications in the algorithm.

## 3. Results and discussion

We begin by analysing the dependence of the average fractal dimension $D$ with three relevant parameters, each corresponding to one major possible direction of anisotropy: longitude, latitude and elevation. Averages with respect to the other two parameters are performed in each case.

In Fig. 2 we summarize the results for the Earth. In order to simplify the notation, we use always $D$ to refer to the fractal dimension irrespective of the parameters that have been averaged. In the top panel of figure 2(a) we show the behaviour of the average fractal dimension versus the longitude; in the middle panel (b), the average is plotted as a function of the latitude; and in the lower panel (c) we plot the average fractal dimension *versus* the elevation. Figure 3 displays the same measures for the Moon. In all these plots, open circles stand for the average value $D$. In order to highlight possible deviations in the distribution of fractal dimensions, we plot, as solid lines, the values $D+\sigma$ and $D-\sigma$, where $\sigma$ is the variance of the measured fractal dimensions. As can be seen, no systematic variations in the distributions are found. A close inspection of the data reveals that, on Earth, the average fractal dimension $D$ is only weakly dependent on longitude (Fig. 2a) or latitude (Fig. 2b) while it shows a complex dependence on elevation (Fig. 2c). Note further the slight change in the vertical scale in the last plot. In contrast, the average fractal dimension $D$ calculated for the Moon (Fig. 3) is always weakly dependent on the three variables: variations are in all cases less than 0.1 around a low value of the fractal dimension $D \approx 1.2$.

As we have already stated, we hypothesize that geometrical properties of topographical features depend in a direct and quantitative manner on the physical processes that shape them. These processes depend on the physical environment (whether the region is embedded in the atmosphere or in the ocean; whether tectonic processes are acting) and on the climatic characteristics of the region (mainly temperature). None of these features varies (on average) as a function of the longitudinal position of the region, and they are only weakly dependent on latitude: there is a temperature gradient towards the Poles and other processes might be affected by the magnetic field of the Earth, which breaks the North-South symmetry. However, the fact that the fractal dimension displays average constant values with respect to longitude and latitude reveals that those gradients have at most a mild effect of geomorphology and supports our hypothesis. In contrast, environment changes systematically with elevation, a variable which appears to play a key role in the characterization of large-scale terrestrial morphology.

In Figure 4a, we show a terrestrial world map with the local fractal dimensions of terrestrial isolines. With the purpose of illustrating graphically the qualitative variation of fractal dimension, we have averaged its value over small cells, such that each coloured area in the plot corresponds now to a region of 0.1° latitude x 0.1° longitude, its colour standing for the average fractal dimension of (isolines contained in) the cell. This procedure amounts to averaging the measures $D(lon,lat,h)$ with respect to a narrow interval of elevations, and yields a fractal dimension depending on two parameters, *lon* and *lat* (once more, for simplicity, we refer to the measure as *D*). For comparison, we summarize in figure 5 the results yielded by performing the same analysis for the Lunar surface. Remarkably, and despite the visible difference in cratering between the two longitudinal hemispheres of the Moon, we obtain a distribution of fractal dimensions almost independent of longitude and latitude (but see below). It is straightforward to conclude that the large-scale, global topography of the Moon is well described by an average fractal dimension between 1.2 and 1.3. These results give further support to the independence of physical processes with longitude and latitude.

The map depicted in Fig. 4a shows several interesting features. Low fractal dimensions (deep blue areas) correspond to smooth regions in oceans: trenches, continental slopes, and continental platforms. These are either regions with steep slopes (trenches and continental slopes) or smooth areas due to the action of processes such as erosion by water and deposition by marine currents (continental platforms). No regions on continents display such low dimensions, with the exception of areas permanently covered by ice (and because elevation measures are taken on top of ice). Fractal dimensions notably increase along the mid-ocean ridge. Moreover, isoline roughness is higher in regions where transform faults are more abundant (North Atlantic, Indian, and South Pacific oceans). Abyssal floors are to be found between the ridge region and the continental platforms, and display a fractal dimension of isolines about 1.3 (dominance of greenish colour). Coastlines signal the boundary between oceanic and continental regions, and are unambiguously identified by our analysis. Fig. 4 (b and c) displays two global isolines, one at 100m below the sea level and a second one at 100m elevation on land. Though the profiles of continents are clearly recognized in both cases, the fractal dimension pinpoints a qualitative change in the landscape through a jump from values close to 1.0 below sea level to notably higher values on continental land. In the latter case, a mild dependence of *D* with latitude along the terrestrial coasts (due to decreases in temperature) is revealed, since areas closer to Polar Regions typically present higher values of *D*. Continents are even more structured, since the effect of latitude and elevation, reflected in strong variations in temperature, is not buffered by the action of oceanic water masses. Furthermore, one expects that water erosion by rivers and ice erosion would increase the complexity of the continental topography.

In order to disentangle the effects of the different processes, we plot in Fig. 6a the fractal dimension as a function of latitude and elevation, averaged over longitude. First, consider the emerged land, i.e. the right portion of the graph. Here two main regions are identified attending to their fractal dimension:

- a yellow-greenish band spanning the region from 40S to 40N and elevations between 0m and 2000m approximately;

- three patches of high roughness (red colour dominant), that correspond, from North to South, to Siberia, to the Himalayan mountains, and to the Andes system (see also Fig. 4a).

Regions of high isoline fractal dimensions closely correspond to locations at a high enough latitude or elevation such that ice is conspicuous all through the year. We propose that the cause of such a rough landscape is the strong erosive action of alpine glaciers. In order to support this hypothesis, we have identified a number of places on Earth where the average yearly temperature is about 0°C (±1°C). White symbols in Fig. 6a show the position of all those locations: they clearly correlate with the position of isolines with the highest fractal dimension. There are two kinds of symbols in the graph. Squares represent actual locations on Earth with yearly average temperature about 0°C. However, not all latitudes have regions with that property, be it because mountains are not high enough at given latitude (as around the equator) or because there are not enough available measures for the location of interest (as in the Andes). In order to fill this gap, we have extrapolated measures taken in locations at the specified latitude but at a too low elevation (hence with average yearly temperatures above 0°C) assuming a decrease of 6°C per 1000m of increase in elevation. This yields a number of extrapolated data represented as circles (climatic data has been obtained from `www.worldclimate.com` and `www.weatherbase.com`).

Above sea level (where the average fractal dimension increases) we identify a range of elevations between 0m and 3000m at almost constant average fractal dimension: this could correspond to the region spanned by terrestrial river systems. The fact that an almost constant fractal dimension characterizes that region is in agreement with the quasi-universality of the topological properties of river basins and river networks [*Hack*, 1957; *Caldarelli et al.*, 2004]. Finally, cold regions with glacial valleys are strongly eroded by ice, which we believe to be responsible for the highest average fractal dimensions observed (above 3000m).

The geometrical properties of regions of Earth below the sea level (left part of Fig. 6a) are more weakly dependent on latitude than continents and coasts, due to the action of large bodies of water that smooth out temperature variations and maintain a more homogeneous environment. In Fig. 6b we plot the fractal dimension as a function of altitude only, i.e. averaged over latitude and longitude. There we see that ocean basins are divided into three main regions: low-dimensional features at high depths –identified with trenches—, rougher structures between -6000m and -3000m –corresponding to the abyssal floor—, and again low-dimensional features between -3000m and coastlines –this region is dominated by continental slopes. Trenches are narrow topographic depressions caused by the subduction of lithospheric plates, thus characterized by steep slopes and consequently rather straight isolines. Abyssal floors are well characterized by a fractal dimension of isolines about 1.3, while continental slopes yield $D$ about 1.1. The continuous increase with depth between regions where those two features dominate is mainly due to the contribution of the mid-ocean ridge, which extends for a length of about 80,000km with a width of 100 to 4,000km. Indeed, the ridge significantly contributes to the topographical features in a range spanning -5000m to almost 0m elevation. Its complex structure systematically heightens the average fractal dimension of isolines as depth increases, since the area of Earth belonging to the ridge province extends with depth. The ocean basin is also decorated with seamounts that contribute at varying depths, mainly from -6000m to -5000m.

The same analysis for the Lunar surface reveals much weaker differences between regions as far as the fractal dimension is concerned (see Fig. 7). Actually, while we know a plethora of different agents simultaneously acting on Earth and tectonism has erased many features from the past, the Moon is a cold, relatively simple planetary body whose surface has been only shaped through meteoritic impacts and the associated lava flows. Two slightly different Lunar regions can be identified only after careful inspection of Fig. 5 and its comparison with the altimetry obtained from the Clementine Project. Several scattered areas with the higher fractal dimensions on the Moon (yellow and orange domains) are found in highlands. These terrains have a rugged relief, with ejecta deposits constituted by large and brecciated rock masses and a plentiful amount of (often superimposed) crater rims. In those regions, the fractal dimension increases locally up to values of 1.4-1.5, but the average keeps around 1.3. Lunar lowlands, having experienced similar processes, are this nonetheless smoothed out by the lava flows, which have washed out part of the previous relief and cause a decrease in the average fractal dimension to a value around 1.2. This is the region of maria (areas of deep blue in Fig. 7), most of them situated on the longitudinal hemisphere facing Earth. The average differences on the Moon are small compared to those measured on the terrestrial surface. Actually, in good agreement with the hypothesis that relates a dominant physical process to an almost constant fractal dimension, a single value of *D*, weakly dependent on elevation, latitude or longitude, characterizes the surface of the Moon: the global Lunar topography is well described by an average fractal dimension around 1.2 in any region (Fig. 3), a narrow interval compared to that of Earth (Fig. 2), where remarkably higher values are attained. Taking impact cratering as the initial (and minimal) action that can be suffered by a planetary body, it seems an immediate consequence that any planet with morphological regions characterized by an average fractal dimension above this value had to be shaped by a *roughening agent* able to increase it. On the contrary, the action of smoothing agents such as lava flows following impacts (as in Lunar maria) should work toward further decreasing *D*.

## 4. Conclusions

We have presented a statistical analysis of the large scale morphology of the surface of Earth and Moon. Quantification of morphological features has been achieved by measuring the local fractal dimensions of isolines. Variations of the average fractal dimension with different parameters have been considered: on Earth, elevation and occasionally latitude appear as key parameters in order to discriminate different observable features.

A more detailed inspection points to a relation between the observed features and the geological processes that shaped the planetary surface. We suggest that the erosive action of a solid (viscous) phase, the erosion of a flowing liquid phase or its action through wave breaking [*Ashton et al.,* 2001; *Sapoval et al.,* 2004], and tectonics, are geological agents which work towards increasing roughness in the landscape. On the other hand, agents causing sediment deposition, formation of pronounced and/or smooth slopes, or erosion through hydrostatic pressure caused by large masses of liquid or ice represent *smoothing agents* able to decrease the characteristic fractal dimension.

We observe a very complex scenario on Earth, according to the broad variation in the local fractal dimension of isolines. These variations correlate to the different specific processes acting in each region. In agreement with our working hypothesis, our analysis identifies a much more uniform Lunar surface, where the only shaping processes are impact cratering and lava flows.

Our analysis complements other studies of the scale invariant properties of terrestrial topology. Studies of the scale invariance of topographic surfaces usually measure correlations in height and perform averages over the horizontal plane [*Gagnon et al.*, 2006]. This procedure washes out angular correlations and cannot yield information on the fractal properties of isolines. It is interesting to remark that both that technique and ours unambiguously establish quantitative differences between continental and oceanic topographies. In [*Gagnon et al.*, 2006], the parameters characterizing the multifractal properties of continents, oceans, and continental margins were clearly different between those regions, but took similar values in separated areas within each of them. Their observations are qualitatively comparable to the acute change in the fractal dimension of isolines that we observe when coasts are crossed. Further, we have observed that oceanic isolines are better represented by mono fractals than continental isolines, where bending of the box-counting curve extends over more scales. Gagnon et al. (2006) ascribe differences between continents and oceans to qualitatively different processes, remarkably the several sources of erosion acting on continents as opposed to the dominant source of erosion (marine currents) in oceans. This is a possible explanation to the systematic deviation from self-affinity that we observe on continents, though the quantitative characterization of natural topography needs (and deserves) further analysis.

We have shown how the relation between the geometrical properties of isolines and the underlying geomorphologic processes gives clues to undertake a systematic classification of global landscape structures and their geological origin. The method is simple and general enough to be applied to any set of planetary topographic data. For instance, an analogous analysis of Mars morphological data could give insights on the presence of an ancient ocean on that planet.

## Acknowledgements

The authors acknowledge discussions with J. Chave, F. Colaiori, D. Fernández-Remolar, J. Ormö, J. Pérez-Mercader, B. Sapoval, A. Vulpiani, and S. Zapperi.

FIGURES

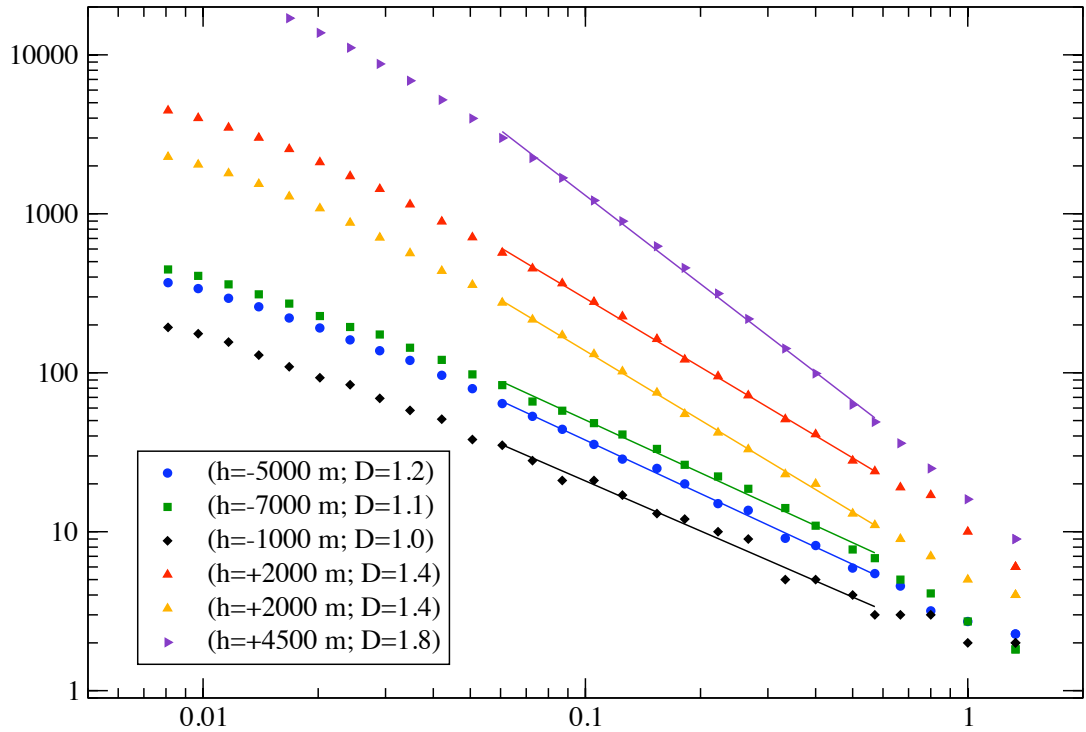

**Figure 1**

Box counting results for isolines at different elevations. Straight lines are interpolations on the range used to calculate *D*. The bending at small scales due to insufficient sampling (not enough points in the isoline close to grid resolution) is seen in all curves, as well as the noisy behaviour at large scales (not enough boxes as they approach the linear size of the cell). There is a clear variation of *D* with elevation. See main text for further explanations.

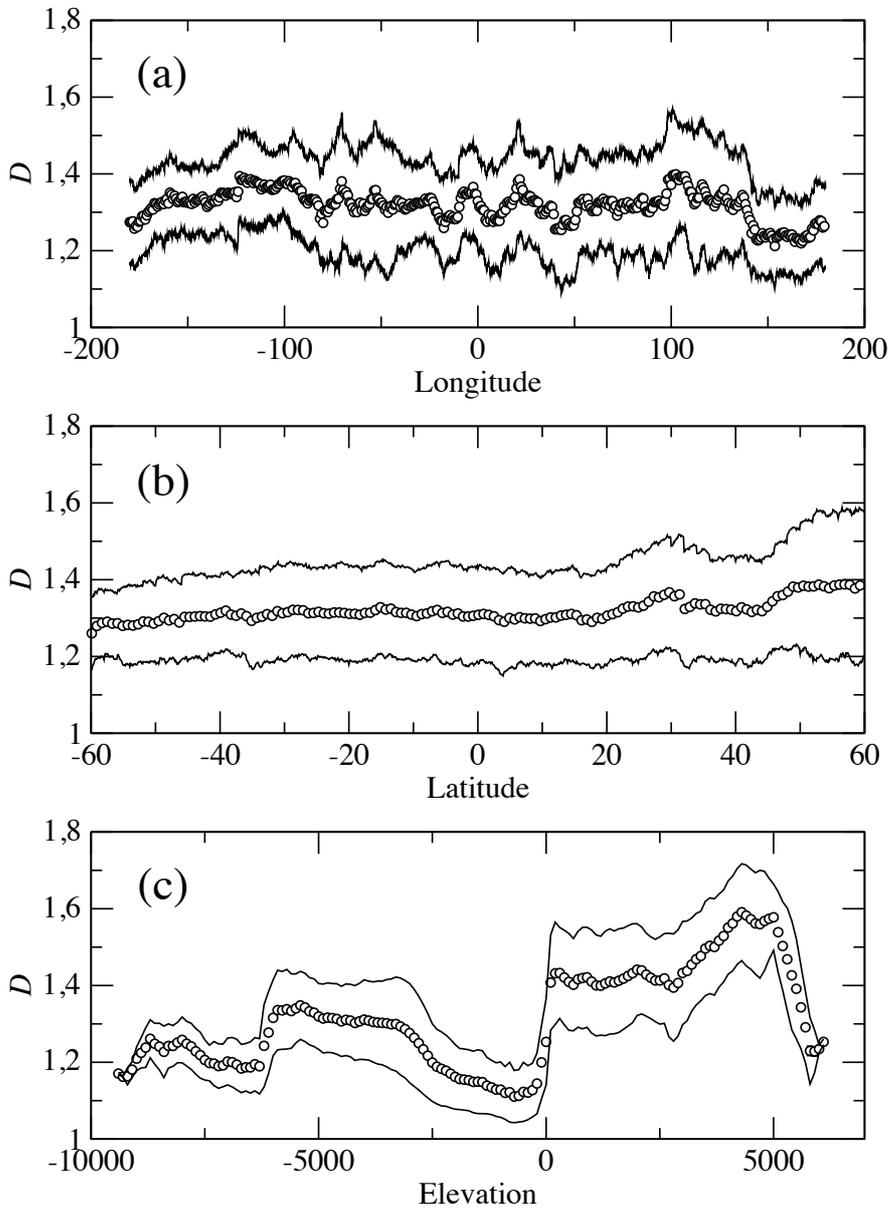

**Figure 2**

Global average fractal dimension $D$ of terrestrial isolines versus a) longitude, b) latitude, c) elevation. In each of the plots $D(lon,lat,h)$ is averaged with respect to two parameters, and plotted against the third, as indicated. Red symbols stand for average values; the statistical error of those measures is smaller than the symbol size. Blue lines represent $D+\sigma$ and $D-\sigma$, where $\sigma$ is the standard deviation of the distribution of fractal dimensions.

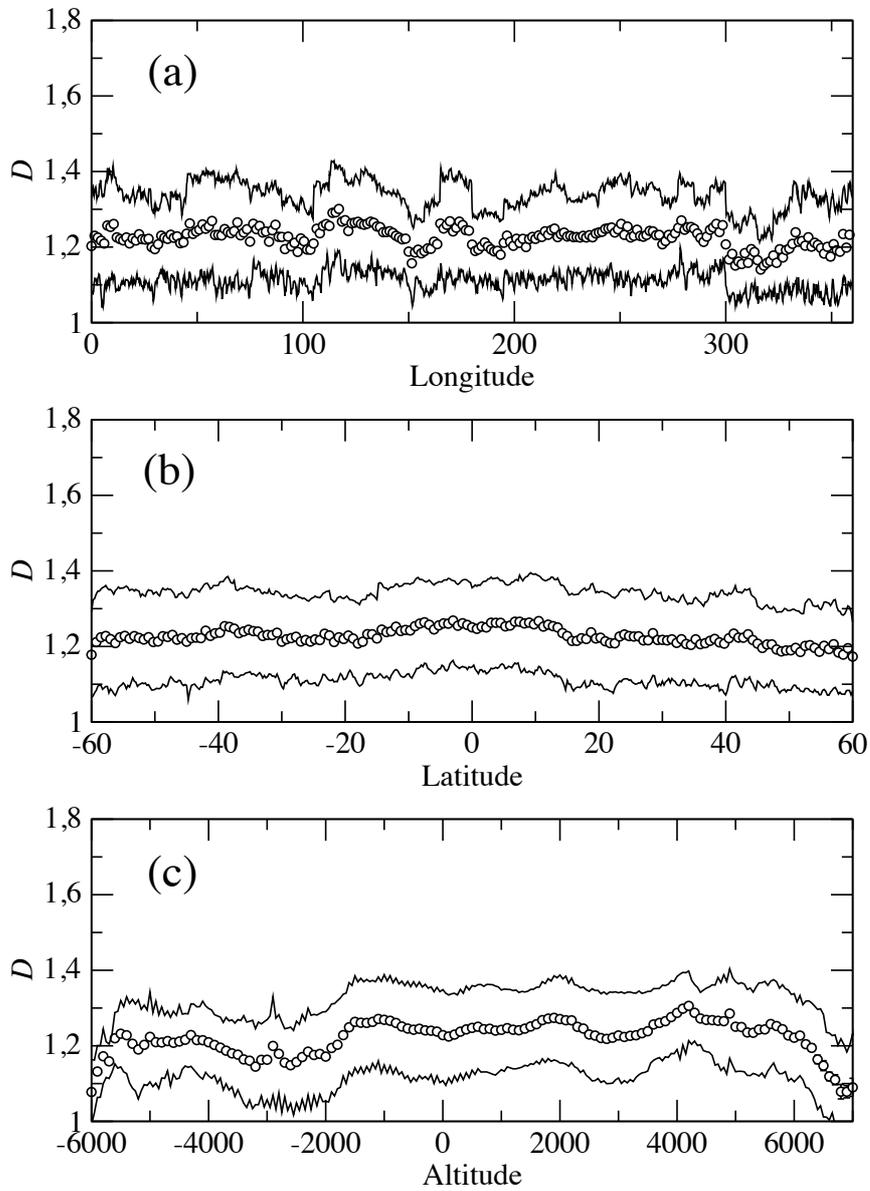

**Figure 3**

Global average fractal dimension *D* of Lunar isolines versus a) longitude; b) latitude, c) elevation. Representation as in Figure 2.

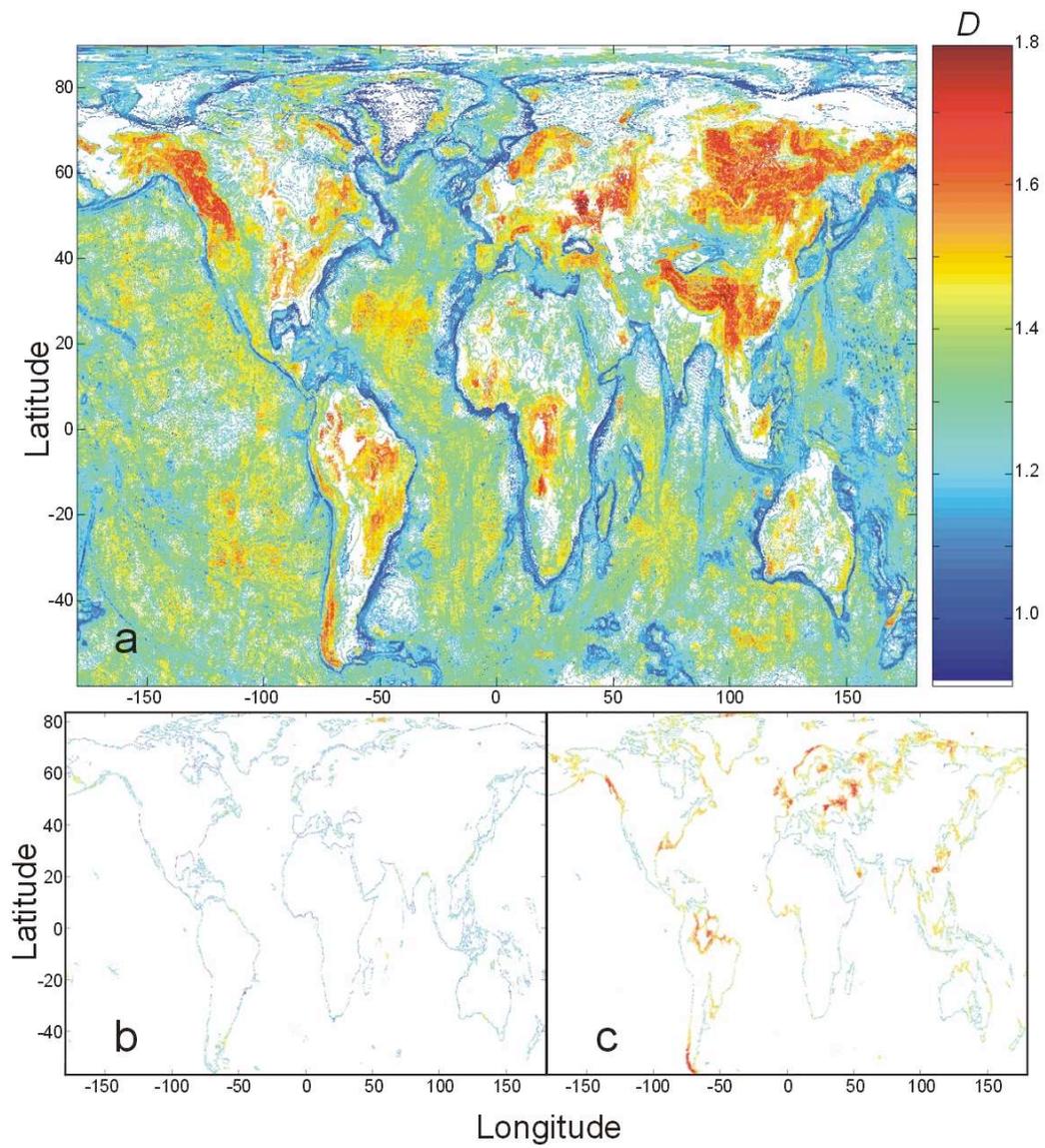

**Figure 4**

(a) World map representing the local fractal dimensions of isolines. Each point of the map corresponds to a region of 0.1° latitude x 0.1° longitude. The colour code indicates the average fractal dimension of isolines within each region. The range of variation of $D$ is broad, from 1.0 to about 1.8. (b) Fractal dimension of isolines at -100m depth and (c) at 100m altitude. Below the sea level (b), the combined smoothing effects of water erosion and sedimentation yield low fractal dimensions. Plot (c) evidences the dramatic increase in average fractal dimension when coastlines are crossed.

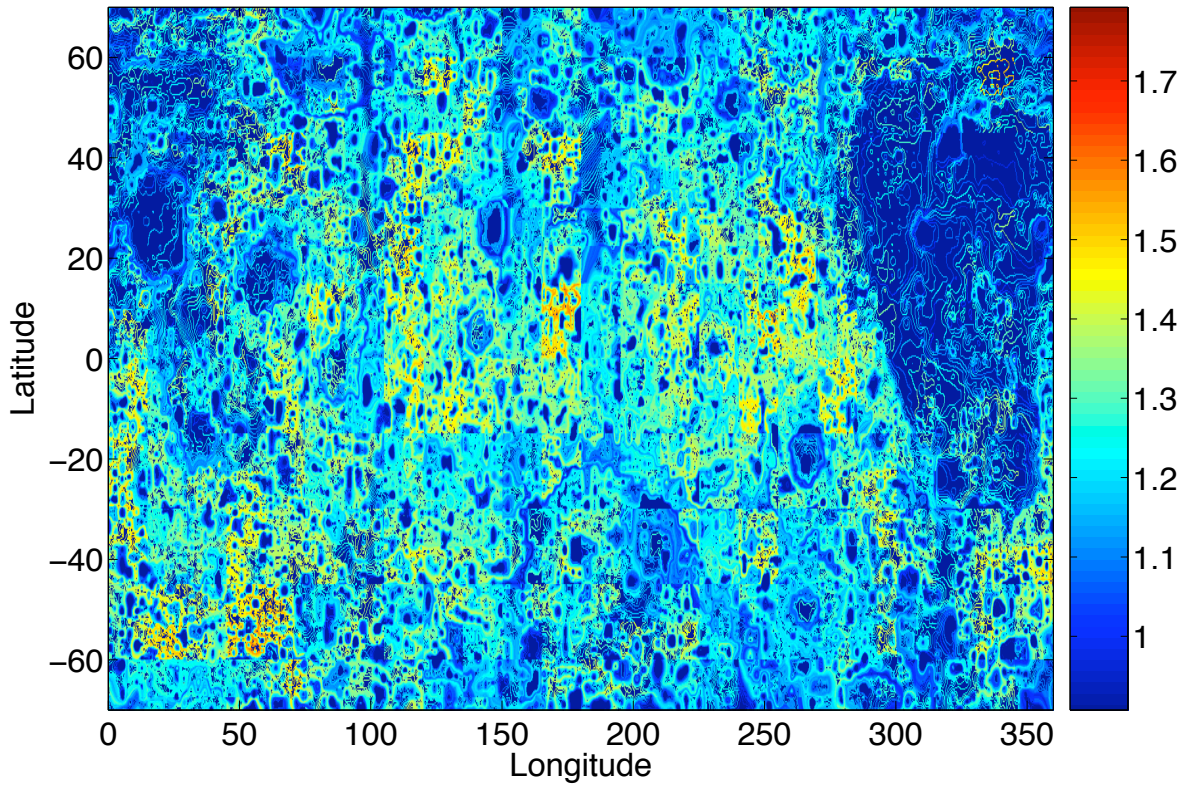

**Figure 5**

Map of the local fractal dimensions of Moon isolines. A weakly varying fractal dimension, with values between 1.2 and 1.3, speaks for the homogeneity of the Lunar surface.

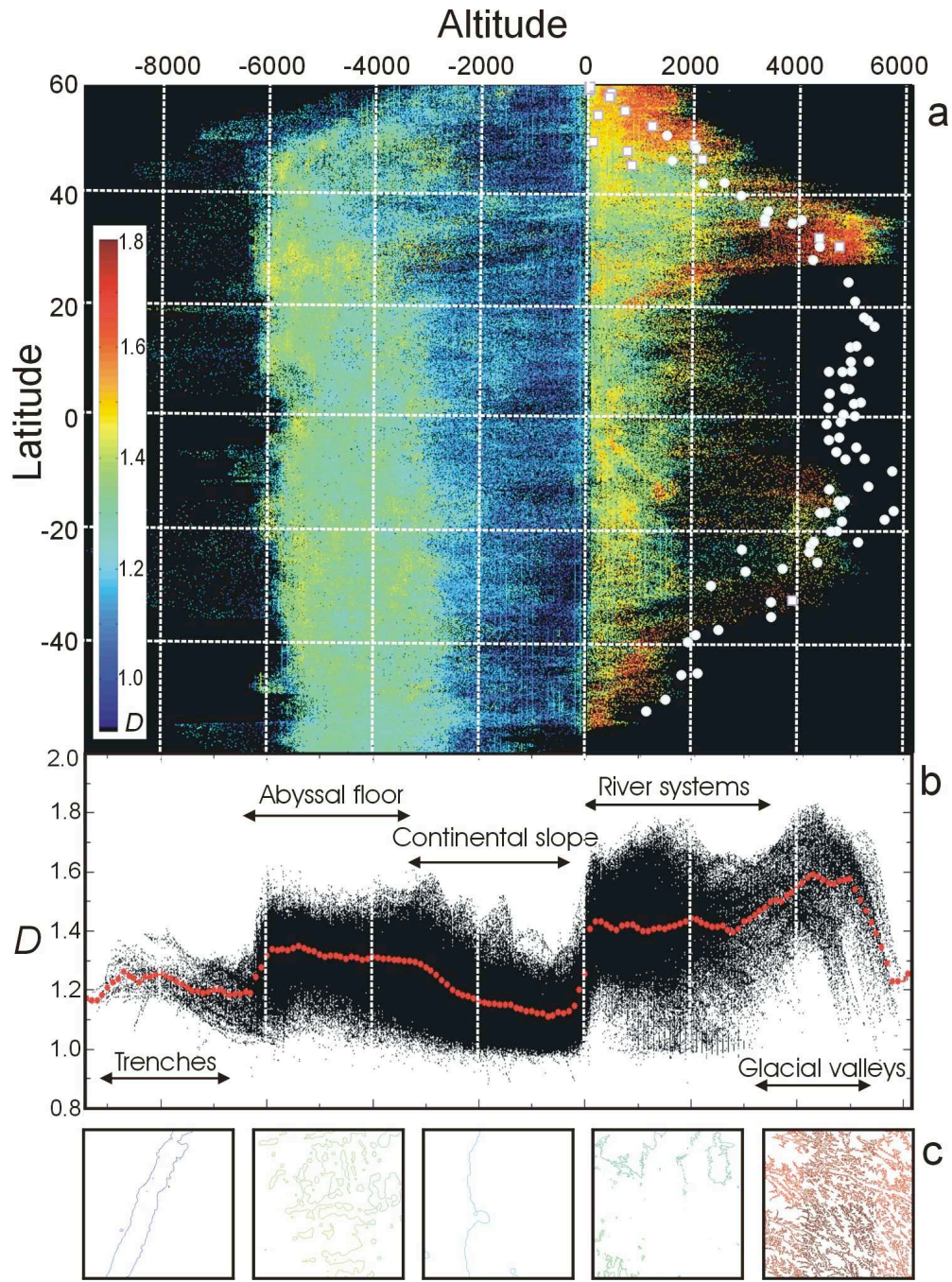

**Figure 6**

(a) Fractal dimension as a function of latitude and elevation (averaged with respect to longitude). Isolines are measured at intervals of 100 meters. Open symbols correspond to locations with yearly average temperature about 0°C (see main text). (b) Fractal dimension *vs* altitude. Black dots correspond to the whole data set represented in colour in (a), while red symbols yield their average value as a function of altitude. This plot contains an information equivalent to that in Fig. 2c, we include it here for better comparison with (a) in this figure. (c) Examples of isolines characterizing major terrestrial geomorphologic features. Each plot covers a surface of 4° latitude x 4° longitude. From left to right, they represent a deep trench (elevation is -7000m, center of plot is at 32S-177W), the abyssal floor (-5000m, 18S-8E), a piece of continental slope in the same plate (-1000m, 18S-12W), a typical continental land dominated by river systems (+2000m, 32N-107W), and the rough contour of high mountains eroded by ice (+4500m, 32N-97E).

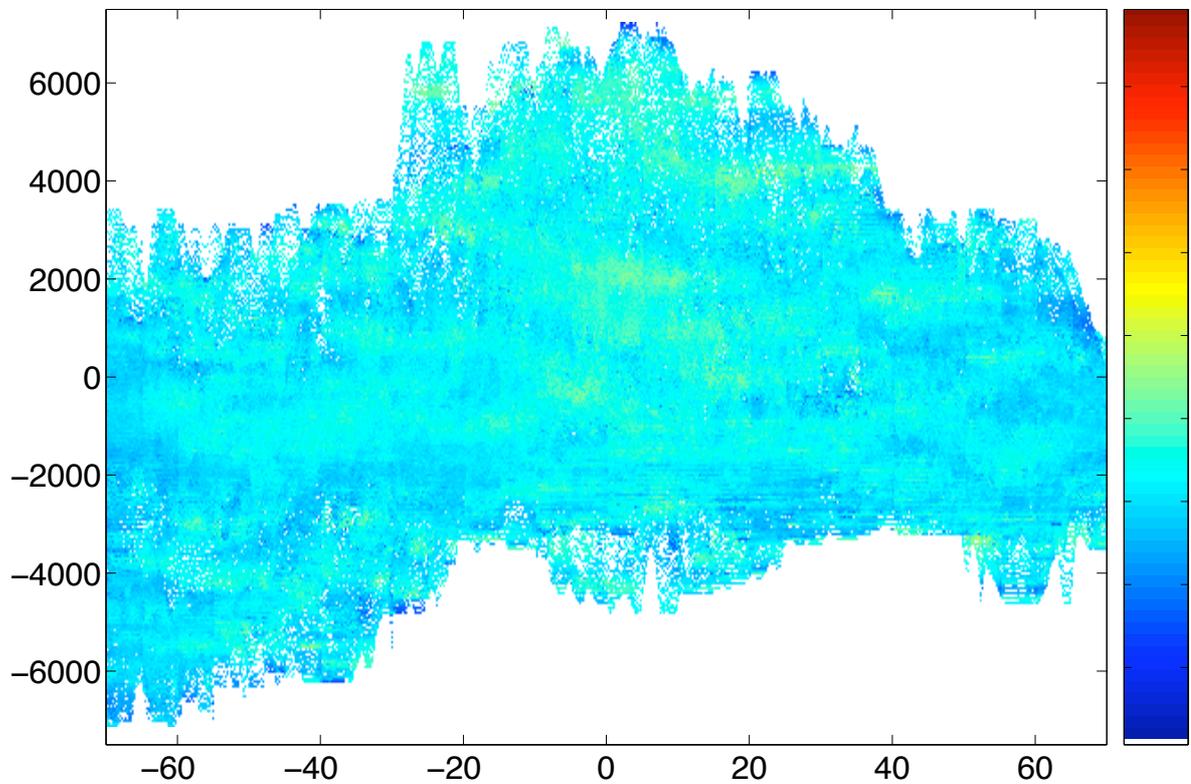

**Figure 7**

Fractal dimension of Moon isolines as a function of latitude and elevation (averaged with respect to longitude). This is analogous to Fig. 6a for Earth.